\documentclass[10pt]{article}
\usepackage[utf8]{inputenc}
\usepackage{graphicx }
\usepackage{subfigure}
\usepackage[colorlinks,citecolor=blue,urlcolor=blue,linkcolor=blue]{hyperref}
\usepackage{authblk}
\usepackage[sort&compress,numbers]{natbib}
\usepackage{amsmath,amssymb}	
\usepackage{empheq}
\usepackage{color,soul}
\usepackage[mathlines]{lineno}

\graphicspath{{./Figs/}}

\title{Buchdahl limit of compact stars in presence of Weyl anomaly}
\author{Waleed El~Hanafy\thanks{waleed.elhanafy@bue.edu.eg}}
\author{G.G.L. Nashed\thanks{nashed@bue.edu.eg}}
\affil{Centre for Theoretical Physics, The British University in Egypt, P.O. Box 43, El~Sherouk City, Cairo 11837, Egypt}

\date{ }

\begin{document}

\maketitle
\begin{abstract}
We setup an anisotropic compact star model in presence of Weyl ``trace" anomaly. We derive an exact interior solution which determines the contribution of the vacuum trace anomaly. We introduce a dimentionless parameter, $\beta$, to characterize this contribution. Applying appropriate matching conditions with the exterior solution, we determine the model parameters in terms of the Weyl anomaly parameter $\beta$ and the compactness parameter, $C=\frac{2GM}{c^2 \mathcal{R}}$ where $M$ and $\mathcal{R}$ are the mass and the radius of the star. We investigate the parameter space $\{\beta, C\}$ and the corresponding modifications of Buchdahl limit on the maximum compactness. We use astrophysical observations of mass and radius of the pulsar PSR J0740+6620 to constrain the Weyl anomaly parameter $\beta$. Also, we investigate the Mass-Radius diagram with other observational constraints from NICER and LIGO/Virgo collaboration.
\end{abstract}

\section{Introduction}\label{Sec:intro}

Neutron star (NS) is extremely dense astrophysical object, with mass about 1.4$M_\odot$, where $M_\odot$ is the solar mass, whereas it is compressed into a sphere only about 10 kilometers in radius. This extreme density creates conditions where both general relativity and quantum mechanics are important. Therefore, quantum gravity might become significant at the NS core. Indeed, quantum gravity is expected to become important at very small scales near the Planck scale, while NSs are not quite at this scale. However, they are much closer to it, at a good approximation, where their cores are at high curvature and energy. This makes NSs the best natural laboratories for studying these effects. It is well known that a full quantum gravity theory is not yet available. Therefore, it is important to investigate possible impact of quantum effects on Einstein field equations at the semi-classical level.

In this sense, we consider the role of quantum effects due to Weyl anomaly on compact star interiors. Weyl anomaly is one of the interesting phenomena in quantum field theory on curved spaces that arises at one-loop level for a collection of conformal fields \cite{Duff:1993wm,Deser:1993yx}. It is known that quantum corrections due to Weyl anomaly are geometric in nature and takes the form of higher-derivatives terms added to Einstein field equations. On cosmic scales, it has been shown that Weyl anomaly could play a vital role to resolve the initial big bang singularity replacing it by a milder finite time singularity of type II \cite{Awad:2015syb}. On the other hand, Weyl anomaly has been adopted to investigate the interior structure of quantum black hole (BH) \cite{Ho:2023eem}.

One of the most important theoretical constraints on compact star models is known as Buchdahl limit. This sets an upper limit on the maximum compactness, $C < 8/9$ (i.e. $M/\mathcal{R}< \frac{4 c^2}{9 G}$) as obtained by \cite{PhysRev.116.1027}. Buchdahl limit is specifically derived for isotropic or mildly anisotropic fluid with spherically symmetric spacetime configuration by applying general relativity. We will call it the classical Buchdahl limit. Therefore, this limit can be easily violated by relaxing one or more of these assumptions. It has been shown that by relaxing mild anisotropy condition, Buchdahl limit is violated for strong anisotropic stars by introducing elasticity for instance, $C\approx 0.924$,  even in GR \cite{Alho:2022bki,Alho:2021sli}, see also \cite{Raposo:2018rjn,Cardoso:2019rvt}. Similarly, the classical Buchdahl limit is violated within Rastall gravity where the fluid maximum compactness reaches $C_\text{max}\approx 0.93$ \cite{ElHanafy:2022kjl,ElHanafy:2023vig}. \textit{The aim of the present study is to derive an exact interior solution of anisotropic compact star in presence of Weyl anomaly. Thus, we investigate possible modification to the classical Buchdahl limit and astrophysical constraints on the Mass-Radius (MR) diagram.}

In organize the paper as follows: In Section \ref{Sec:Weyl}, we derive the field equations of a compact star interior in presence of Weyl anomaly. In Section \ref{Sec:Astro_constr}, we investigate the modified Buchdahl limit due to Weyl anomaly. We use astrophysical constraints on the mass and the radius of the pulsar PSR J0740+6620 to determine the model parameters, while the MR diagram has been illustrated along with astrophysical constraints from NICER, LIGO and Virgo observation. In Section \ref{Sec:summary}, we summarize the work.


\section{Compact star model in presence of Weyl anomaly}\label{Sec:Weyl}

We assume the spacetime configuration of the star interior is described by a static spherically symmetric metric in Schwarzschild coordinates
\begin{equation}\label{eq:metric}
    ds^2=e^{\nu(r)}c^2 dt^2 - e^{\lambda(r)}dr^2 -r^2d\theta^2 -r^2 \sin{\theta}^2 d\phi^2.
\end{equation}
We write the semi-classical Einstein equations, where gravity is treated classically and its source is represented by the expectation value in some quantum state of the matter stress-energy tensor operator
\begin{equation}\label{eq:Ein_feqns}
    \mathcal{G}_{ab}=R_{ab}-\frac{1}{2} g_{ab} R=\kappa\langle T_{ab}\rangle,
\end{equation}
where $\kappa=8\pi G_N/c^4$ with $G_N$ being the Newtonian constant, $c$ is the speed of light, and the expectation value of the effective stress-energy tensor $\langle T_{ab}\rangle$ related to quantum loops. It is well known that the expectation value of the trace of the energy-momentum tensor, $\langle T \rangle \equiv \langle T{^a}{_a}\rangle$, vanishes at the classical level for theories which hold conformal symmetry. A typical case is given by massless quantum chromodynamics (QCD). Conformal symmetry, however, is broken at the quantum level. This violation is quantified via the trace anomaly, which has the anomalous term proportional to the gluon condensate due to the running of the strong coupling constant.

\textit{The trace anomaly}. It has been shown that, at finite temperature $\Theta$ and baryon chemical potential $\mu_B$, the condensate should depend on $\Theta$ and $\mu_B$ and we can decompose the trace anomaly into two parts {\cite{Fujimoto:2022ohj}}, $\langle T \rangle=\langle T \rangle_{\Theta,\mu_{B}} + \langle T \rangle_{0}$,
where $\langle T \rangle_{\Theta,\mu_{B}}$ denotes the matter contribution to the trace anomaly and $\langle T \rangle_{0}$ denotes the contribution of the vacuum expectation value (VeV) at $\Theta=0$ and $\mu_{B}=0$. In Ref. {\cite{Fujimoto:2022ohj}}, only the matter part has been discussed, where the trace anomaly has been expressed in terms of thermodynamic quantities. This gives $\langle T \rangle_{\Theta,\mu_{B}} =\varepsilon-3P$, where $\varepsilon$ and $P$ denote the energy density and the isotropic pressure inside a compact stellar object. It has been shown that the sound velocity peaks up, above the conformal limit ($v_s^2=c^2/3$) at density few times the nuclear saturation energy density ($\varepsilon_\text{sat}=150$~MeV/fm$^3$). This has been attributed to the derivative contribution from the trace anomaly that steeply approaches the conformal limit.

In the present study, we assume a more general case of anisotropic pressure, i.e.
\begin{equation}\label{eq:matt_trace}
    \langle T \rangle_{\Theta,\mu_B}=\rho c^2-p_r-2 p_t,
\end{equation}
where $p_r$ and $p_t$ denote the radial and the tangential pressures.
On the other hand, in four dimensional spacetime, one loop quantum correction leads to a trace anomaly of the stress-energy tensor of conformal field theory. The VeV trace anomaly can be written as {\cite{Birrell:1982ix,Duff:1993wm,Deser:1993yx}}
\begin{equation}\label{eq:Weyl_VeV}
    \langle T \rangle_{0}=\langle T{^a}{_a}\rangle_{0}=\tilde{\alpha} \left(I_4- \frac{2}{3}\Box R \right)+\tilde{\beta} E_4,
\end{equation}
where $I_4$ denotes the square of Weyl tensor (known as Type B anomaly) and $E_4$ denotes Euler density which is the Gauss-Bonnet scalar (known as Type A anomaly).
The two coefficients $\tilde{\alpha}$ and $ \tilde{\beta}$ are regularization-scheme-independent and they can be given by $\tilde{\alpha} =\frac{1}{120(4\pi)^2}\left(n_{s}+6 n_{f}+12 n_{v}\right)$ and $\tilde{\beta} =-\frac{1}{360(4\pi)^2}\left(n_{s}+11 n_{f}+62 n_{v}\right)$,
where $n_{s}$ is the number of scalars, $n_{f}$ the number of Dirac fermion and $n_{v}$ is the number of vector fields \cite{Duff:1993wm,Deser:1993yx,Birrell:1982ix}. Notably, the contributions of $n_{s}$, $n_{f}$ and $n_{v}$ show that $\alpha$ is positive and $\beta$ is negative. We note that the trace anomaly {\eqref{eq:Weyl_VeV}} is valid for theories where the matter field is classically conformally invariant \cite{Duff:1993wm}. Otherwise, due to lack of conformal invariance, the anomalies still survive, but expected to be accompanied by contributions to $g^{ab}\langle T_{ab} \rangle$\footnote{The anomaly in the theories, which are not conformally invariant, exists due to non-commutation of regularization and taking the trace operations. Therefore, it is defined as Anomaly=$g^{ab}\langle T_{ab} \rangle_{reg} -\langle g^{ab} T_{ab} \rangle_{reg}$, where the second term has to vanish in conformal invariance limit \cite{Duff:1993wm}.}. In the present work, we drop these corrections. As we have mentioned earlier in this section that the matter contribution to the trace anomaly has been already studied inside compact stellar objects {\cite{Fujimoto:2022ohj}}, where the trace of the matter field does not vanish. In the present work, we study possible additional contribution of the VeV, i.e. $\langle T \rangle_0$ as given by {\eqref{eq:Weyl_VeV}}, to the stress-energy tensor.

\textit{The field equations}. The effective stress-energy tensor, in the field equations {\eqref{eq:Ein_feqns}}, clearly must satisfy the covariant conservation law $\nabla_a \langle T{^a}{_b} \rangle=0$. However, additional constraints should be imposed to fully determine the energy-stress tensor \cite{Christensen:1977jc}, see also \cite{Cai:2009ua,Cai:2014jea}. In the present study, we consider the case where the matter fields coupled to a conformally flat metric, i.e. $g_{ab} = \Omega^2(x^a) \eta_{ab}$, which requires the vanishing of Weyl tensor. We note that the metric \eqref{eq:metric} clearly does not fulfill the conformally flat form, since we choose to work with Schwarzschild coordinates which provide a good frame for direct measurements of real astrophysical objects. However, the vanishing of the Weyl tensor condition guarantees that the metric is conformally flat. Since all equations are invariant under the diffeomorphism group, we should not be worried about this issue. This leaves only Euler density contribution to the vacuum trace anomaly \eqref{eq:Weyl_VeV} in addition to a local derivative term. In this case, the renormalized VeV of the stress-tensor trace reduces to \cite{Birrell:1982ix,Duff:1993wm,Awad:2015syb}
\begin{equation}\label{eq:confflat_TA}
    \langle T{^a}{_a}\rangle_{0}^{\text{ren}}=-6\alpha \Box R+ \beta \left[\frac{1}{3}R^2-R^{ab}R_{ab}\right],
\end{equation}
where $\alpha=-\tilde{\alpha}/144$ and $\beta=\tilde{\beta}/8$, see \cite{Birrell:1982ix}. Since $\langle T{^a}{_a}\rangle_{0}^{\text{ren}}$ is derived from effective action, it must covariantly conserved $\nabla_a \langle T{^a}{_b} \rangle_{0}^{\text{ren}}=0$. Notably,  $\alpha$ is a coefficient of a total derivative term within the anomaly, then one might select a regularization scheme where $\alpha$ equals zero or to incorporate a local counter term to eliminate this term. In the following, we use the former regularization scheme by setting $\alpha=0$ and also keep in mind that $\beta>0$ is excluded.

In addition to the geometric tensors in equation {\eqref{eq:confflat_TA}}, one might consider the contribution of another local and conserved (non-geometric) tensor, that is, $T_{ab}$, the matter stress-energy tensor \cite{Birrell:1982ix}. Thus, the field equations {\eqref{eq:Ein_feqns}} read\footnote{In order to have $\beta$ dimensionless, we use the transformation $\beta \to \frac{\beta \ell^2}{\kappa}$, where $\ell$ is a length scale associated to the mass distribution of the compact object. It proves convenient to choose it as the star radius $\mathcal{R}$.}
\begin{equation}\label{eq:Field_Eqns}
    R_{ab}-\frac{1}{2} g_{ab} R-\beta \left[\frac{1}{12}g_{ab} R^2-R^{cd}R_{cadb}\right]=\kappa T_{ab}.
\end{equation}
It is to be noted that quantum corrections due to Weyl anomaly (associated to the parameter $\beta$) are fully geometric and represented by higher derivatives terms added to Einstein field equations. Accidentally, this term is covariantly conserved in conformally flat spacetimes {\cite{Birrell:1982ix}}, it is not conserved otherwise.

\textit{The model assumptions and the interior solution}. We note that by applying the field equations {\eqref{eq:Field_Eqns}} to the spacetime {\eqref{eq:metric}} with a matter field {\eqref{eq:matt_trace}}, one obtains a system of three differential equations in five unknowns. In principle, two conditions must be imposed to close the system. Usually, two equations of state are needed, $p_r (\rho)$ and $p_t (\rho)$, to solve the system. In the present model, we impose two conditions to close the system: (1) The conformal flatness condition, i.e. vanishing of Weyl tensor. (2) The vacuum trace anomaly exactly cancels the matter trace anomaly, which leads to trace free effective stress-energy tensor. Explicitly, we require
\begin{equation}\label{eq:model_assumptions}
C^{\mu\nu\lambda\sigma}=0,\, \quad \text{and}\,  \quad \langle T \rangle=\langle T \rangle_{\Theta,\mu_{B}} + \langle T \rangle_{0}=0.
\end{equation}
We note that the second assumption derives the Einstein tensor trace to vanish, $\mathcal{G}=0$, and consequently one obtains $R=0$ inside the stellar object. In absence of Weyl anomaly, the traceless energy-stress tensor of matter sector could be a good candidate to describe compact stars. For example, quark matter exhibits an approximate conformal symmetry, where the trace of the energy-stress tensor fulfills $\langle T \rangle_{\Theta,\mu_B} \to 0$, see \cite{Annala:2023cwx}. If the conjectured quark matter is the true ground state of matter at zero pressure as proposed by Witten \cite{Witten:1984rs} (see also \citep{Farhi:1984qu}), the quark star model could be the real model of all pulsars in fact. Another example in terms of scalar-tensor theories where 136 equation of state (EoS) have been imposed \cite{Podkowka:2018gib}. It has been shown that the MR curves determine a compactness, for which $\langle T \rangle_{\Theta,\mu_B} = 0$, at the stellar center is $C = 2G M/(\mathcal{R} c^2) \sim 0.524$ approximately EoS independent.

By imposing the model assumptions \eqref{eq:model_assumptions} into the field equations \eqref{eq:Field_Eqns}, we obtain the following exact solution
\begin{equation}\label{eq:model_solns}
\nu=\ln\left[\frac{1}{4}C_1 r^2 \sec^2 \chi\right],\quad \text{and}\quad \lambda-\nu=2\ln C_2.
\end{equation}
where $\chi\equiv \frac{1}{2}\sqrt{C_1}(C_2 r- C_3)$, and $C_1$, $C_2$ and $C_3$ are arbitrary constants.  It proves convenient to write the metric coefficients \eqref{eq:metric} in terms of dimentionless parameters. Therefore, we write
\begin{equation}
    C_1 = a_0^2/ \mathcal{R}^2 , C_2= a_2, C_3 = - a_1 \mathcal{R},
\end{equation}
where the newly introduced dimensionless parameters are $a_0$, $a_1$ and $a_2$. Thus, the metric becomes
\begin{equation}
    g_{ab}=diag\left(\frac{a_0^2 x^2}{4\cos^2{\chi}},-\frac{a_0^2 a_2^2 x^2}{4\cos^2{\chi}}, -r^2, -r^2 \sin^2{\theta}\right),
\end{equation}
where $\chi=-\frac{1}{2}a_0(a_1+a_2 x)$ and $x=r/\mathcal{R}$. We also define a critical density
$\rho_\text{crit}\equiv\frac{1}{\kappa c^2 \mathcal{R}^2}$, then we write the dimensionless density and dimensionless pressures
\begin{eqnarray}
\nonumber    \bar{\rho}&=&\frac{\rho}{\rho_\text{crit}}=\frac{4\cos^{2}{\chi}+a_0^2a_2^2 x^2+4 a_0 a_2 x \sin{\chi}\cos{\chi}}{a_0^2 a_2^2 x^4}+\frac{\beta}{a_0^4 a_2^4 x^8} \left[ \right. 80 \cos^4{\chi}-24 a_0^2 a_2^2 x^2 \cos^2{\chi}+ a_0^4 a_2^4 x^4\\
    &&+32 a_0 a_2 x \sin{\chi}\cos^3{\chi}-8 a_0^3 a_2^3 x^3 \sin{\chi} \cos{\chi} \left.\right],\label{eq:densx}\\[5pt]
\nonumber    \bar{p}_r&=& \frac{p_r}{\rho_\text{crit}c^2}=\frac{4 a_0 a_2 x \sin{\chi} \cos{\chi}+12 \cos^2{\chi}-a_0^2 a_2^2 x^2}{a_0^2 a_2^2 x^4}+\frac{\beta}{a_0^4 a_2^4 x^8} \left[ \right. 32 a_0 a_2 x \sin{\chi} \cos^3{\chi}+48 \cos^4{\chi}\\
    &&-8 a_0^3 a_2 ^3 x^3 \sin{\chi} \cos{\chi} - a_0^4 a_2^4 x^4 -8 a_0^2 a_2^2 x^2 \cos^2{\chi}\left.\right],\label{eq:rpressx}\\[5pt]
\nonumber    \bar{p}_t&=& \frac{p_t}{\rho_\text{crit}c^2}=\frac{a_0^2 a_2^2 x^2 - 4 \cos^2{\chi}}{a_0^2 a_2^2 x^4}-\frac{\beta}{a_0^4 a_2^4 x^8} \left[ \right. 8 \cos^4{\chi}-16 a_0^2 a_2^2 x^2 \cos^4{\chi}+8 a_0^2 a_2^2 x^2 \cos{\chi^2}+a_0^4 a_2^4 x^4\\
    &&+ 64 a_0 a_2 x \sin{\chi} \cos^3{\chi} \left.\right].\label{eq:tpressx}
\end{eqnarray}

\textit{The matching conditions}. It has been shown that exact exterior BH vacuum solutions can be found by solving the semi-classical Einstein field equations with Weyl anomaly. The set of solutions contains two constants of integration which can be interpreted as: mass and $U(1)$ conserved charge of some classical conformal field theory. In principle, the solutions have two branches, one is asymptotically flat and the other asymptotically de Sitter \cite{Cai:2009ua}, see also \cite{Calza:2022szy}, where the latter has been argued to be unstable. Thus, the BH exterior solution can be written as $g_{tt}=1/g_{rr}=1-\frac{2GM}{c^2 r}$, at $r>\mathcal{R}$, in absence of conformal charges \cite{Cai:2009ua}. Thus, we take the following boundary conditions at the stellar surface $r=\mathcal{R}$:
\begin{equation}\label{eq:boundary_conds}
    g_{tt}=1-C, \quad g_{rr}=\frac{1}{1-C}, \quad \text{ and } \quad p_r(\mathcal{R})=0.
\end{equation}
Consequently, we determine two sets of the dimensionless model parameters, explicitly in terms of $\beta$ and $C$, as follows: The first set is
\begin{eqnarray}
    a_0 &=&-\frac{\sqrt{\zeta}}{2\beta C-1},\quad    a_1 =\frac{\left[2(1-C)(2\beta C -1) \arccos{\left(\frac{\sqrt{\zeta}}{2(2\beta C -1)\sqrt{1-C}}\right)-\sqrt{\zeta}}\right]}{(1-C)\sqrt{\zeta}},\quad
    a_2 =\frac{-1}{1-C}.\label{eq:const1}
\end{eqnarray}
The second set is
\begin{eqnarray}
   a_0= -\frac{\sqrt{\zeta}}{2\beta C-1},\quad
    a_1 = \frac{\left[2(1-C)(2\beta C -1) \left(2\pi-\arccos{\left(\frac{\sqrt{\zeta}}{2(2\beta C -1)\sqrt{1-C}}\right)}\right)-\sqrt{\zeta}\right]}{(1-C)\sqrt{\zeta}},\quad
    a_2 = \frac{-1}{1-C},\label{eq:const2}
\end{eqnarray}
where $\zeta=-C\left[(9C-8)(\beta C-1)^2+4\beta C\right]$. Clearly, all parameters can be expressed in terms of the star compactness $C$, as in the GR case, in addition to the Weyl anomaly parameter $\beta$.

We define the mass function
\begin{equation}
     m(r)=4\pi \int_0^r \rho(r') r'^2 dr'.
\end{equation}
Substituting from \eqref{eq:densx}, we write mass distribution inside the stellar object in Weyl anomaly model
\begin{equation}
    m(x)=\frac{4\pi \mathcal{R} (a_0^2 a_2^2 x^2 - 2 \cos{\chi}- 2)}{a_0^2 a_2^2 c^2 \kappa x}-\frac{4\pi \mathcal{R} \beta}{a_0^4 a_2^4 c^2 \kappa x^5}\left(a_0^2 a_2^2 x^2 - 2 \cos{\chi}- 2\right)^2.
\end{equation}
Imposing the boundary condition $m(r\to 0)=0$ sets $a_0=\pm \pi/a_1$. Otherwise, the object is a BH. Thus, the final mass function form reads
\begin{equation}\label{eq:mass_fn}
    m(x)=\frac{4 \mathcal{R} [\pi^2 a_2^2 x^2 - 2 a_1^2 (\cos{\chi}- 1)]}{\pi a_2^2 c^2 \kappa x}-\frac{4 \mathcal{R} \beta}{\pi^3 a_2^4 c^2 \kappa x^5}\left[\pi^2 a_2^2 x^2 - 2 a_1^2 (\cos{\chi}- 1)\right]^2.
\end{equation}
Astrophysical observations on the masses and radii of compact objects set a direct constraint on the Weyl anomaly parameter $\beta$. This will be seen shortly in the next section.

\textit{The hydrostatic equilibrium equation}. We note that the VeV trace anomaly \eqref{eq:confflat_TA} reads
\begin{equation}\label{eq:Wtrace2}
    \langle T \rangle_0^\text{ren}= \frac{-\beta \mathcal{R}^2}{C_1^{5/2}C_2^4 r^8 \kappa}\left[C_1^{5/2} C_2^4 r^4-8 \cos{\chi}^3\left(\sqrt{C_1}\cos{\chi}(C_1 C_2^2 r^2-6)+4C_1 C_2 r \sin{\chi}\right)\right].
\end{equation}
It is straight forward to verify that $\langle T \rangle=0$ as $\langle T \rangle_{\Theta,\mu_B}=-\langle T \rangle_0^\text{ren} \neq0$, in addition to the conservation equations $\nabla_a \langle T{^a}{_b} \rangle_{\Theta,\mu_B}=0$ and $\nabla_a \langle T{^a}{_b} \rangle_{0}^\text{ren}=0$. Consequently, the effective stress-energy tensor fulfills continuity equation $\nabla_a \langle T{^a}{_b} \rangle=0$. The latter constraint results to the Tolman–Oppenheimer–Volkoff (TOV) equation of hydrostatic equilibrium
\begin{equation}\label{eq:TOV}
 p'_{r}=-\frac{1}{2}(\rho c^2+p_r)\nu'+\frac{2}{r}(p_t-p_r),
\end{equation}
where $'\equiv d/dr$. In Equation {\eqref{eq:TOV}}, the left hand side represents the a repulsive force due to hydrostatic radial pressure $p'_r=-F_h$, while its right hand side includes two forces: The first term, $F_g=-\frac{1}{2}(\rho c^2+p_r)\nu'$, represents relativistic (attractive) gravitational force, the second term, $F_a=\frac{2}{r}(p_t-p_r)$, is due to anisotropy. We note that matter and vaccum are assumed to be minimally coupled, where each sector must individually satisfy the conservation law. In this case, the Weyl anomaly contribution to TOV equation should vanish as $\nabla_a \langle T{^a}{_b} \rangle_{0}=0$. Otherwise, nonminimal coupling between matter and Weyl anomaly (geometric origin) must be assumed, where $\nabla_a \langle T{^a}{_b} \rangle_{\Theta,\mu_B}\neq 0$ and $\nabla_a \langle T{^a}{_b} \rangle_{0}^\text{ren} \neq 0$ but $\nabla_a \langle T{^a}{_b} \rangle=0$. This is similar to quadratic Rastall gravity \cite{ElHanafy:2024cti}.

\section{Buchdahl Limit and Observational Constraints}\label{Sec:Astro_constr}
Since the model constants can be written in terms of the compactness of the stellar object, $C$, and the parameter $\beta$ which identifies the vacuum trace anomaly, the parameter space is $\{\beta, C\}$. In this sense, we investigate possible modifications on the Buchdahl limit due to Weyl anomaly. Therefore, we check the case $p_r(r=0)\to \pm \infty$, which defines Buchdahl limit of the maximum allowed compactness. By setting $a_0=\pm \pi/a_1$, as required to verify that $m(0)=0$, in the radial pressure equation \eqref{eq:rpressx}, Buchdahl limit can be described by the inequality $\zeta>0$. Explicitly, we write
\begin{equation}\label{eq:Buchdahl_ineq}
    -C\left[(9C-8)(\beta C-1)^2+4\beta C\right]>0.
\end{equation}
\begin{figure*}[h!]
    \centering
    \subfigure[The parameter space $\{\beta, C\}$]{\label{fig:fig1a}\includegraphics[width=0.35\linewidth]{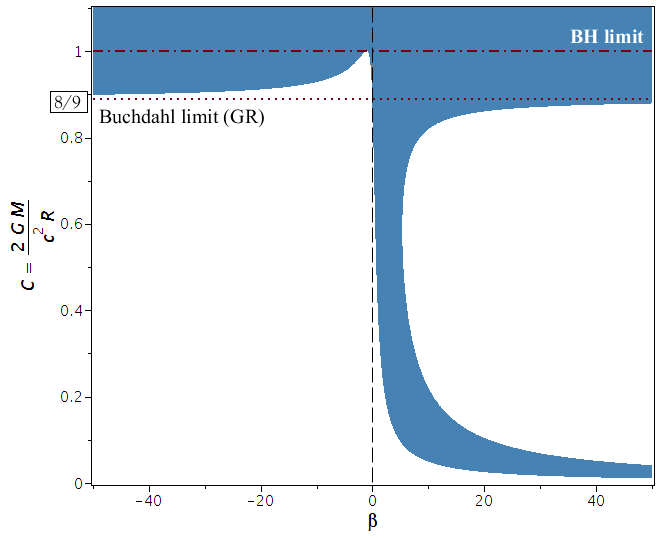}}\hspace{1.5cm}
    \subfigure[Maximum compactness]{\label{fig:fig1b}\includegraphics[width=0.35\linewidth]{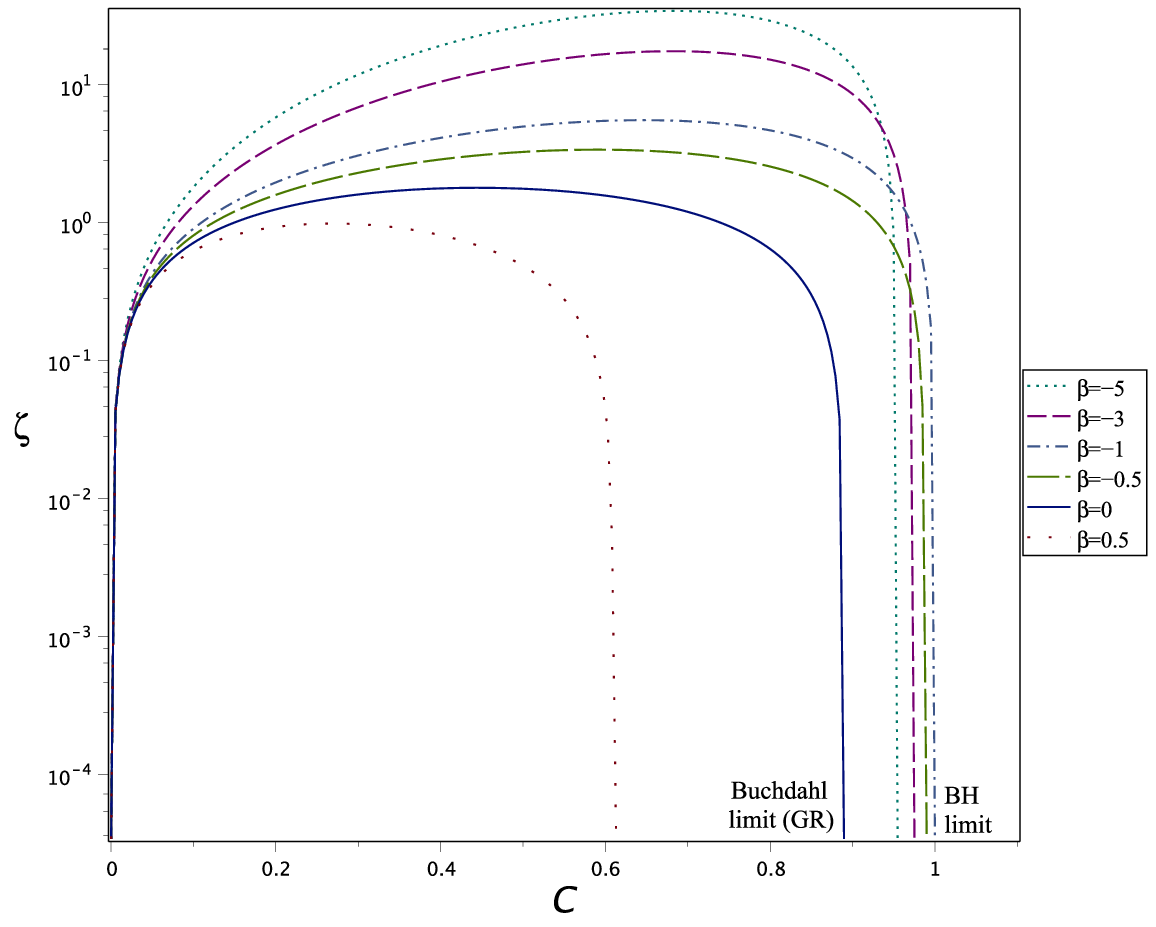}}
    \caption{\subref{fig:fig1a} The parameter space $\{\beta, C\}$ associated to Buchdahl limit, $p_r(0)\to \pm \infty$, of Weyl anomaly. For the GR case ($\beta=0$), the maximum compactness $C<\frac{8}{9}$ as indicated by the dotted horizontal line. For $\beta\neq 0$ the maximum compactness is given by the upper bound of the unshaded regions. Notably the maximum compactness reaches the BH limit (dash-dotted horizontal line) $C\to 1$ as $\beta\to -1$. At large $|\beta|$, the classical GR Buchdahl limit (dotted horizontal line) is recovered, but $\beta>0$ region is excluded. \subref{fig:fig1b} Buchdahl limit of the stellar compactness, at which $p_r$ diverges, is determined by the inequality $\zeta>0$, namely \eqref{eq:Buchdahl_ineq}. At $\beta=0$ the maximum compactness of the GR theory $C\to 8/9$ is recovered. At $\beta\to -1$, the maximum compactness reaches the BH limit $C\to 1$. At large $|\beta|$ in the negative region, Buchdahl limit reduces to the GR limit $C\to 8/9^+$.}
    \label{fig:Fig1}
\end{figure*}
Clearly the GR limit ($\beta=0$) can be recovered, where classical Buchdahl limit is $C<8/9$. Otherwise, the solution of the above inequality is given by the phase space $\{(\beta, C)\}$ as in Figure \ref{fig:fig1a} where the shaded regions are exclude by modified Buchdahl limit and BH limit. We note that positive $\beta$ values are excluded. Remarkably, Buchdahl compactness reaches the BH limit $C\to 1$ as $\beta\to -1$, where the causality constraint at the center $r\to 0$ is fulfilled, i.e. the sound speed, $c_s$, in the radial direction relative to the speed of light, $c$, is $c_s^2/c^2=dp_r/d\rho\to 1$. At large $\beta$ (negative values), the GR Buchdahl constraint on the maximum compactness is recovered where $C\to 8/9$ from above. In Figure \ref{fig:fig1b}, we plot $\zeta$ verses $C$ at different values of $\beta$. It shows that $\zeta>0$ curve defines the maximum compactness at $\zeta\to 0^+$.

\begin{figure}
    \centering
    \includegraphics[width=0.36\linewidth]{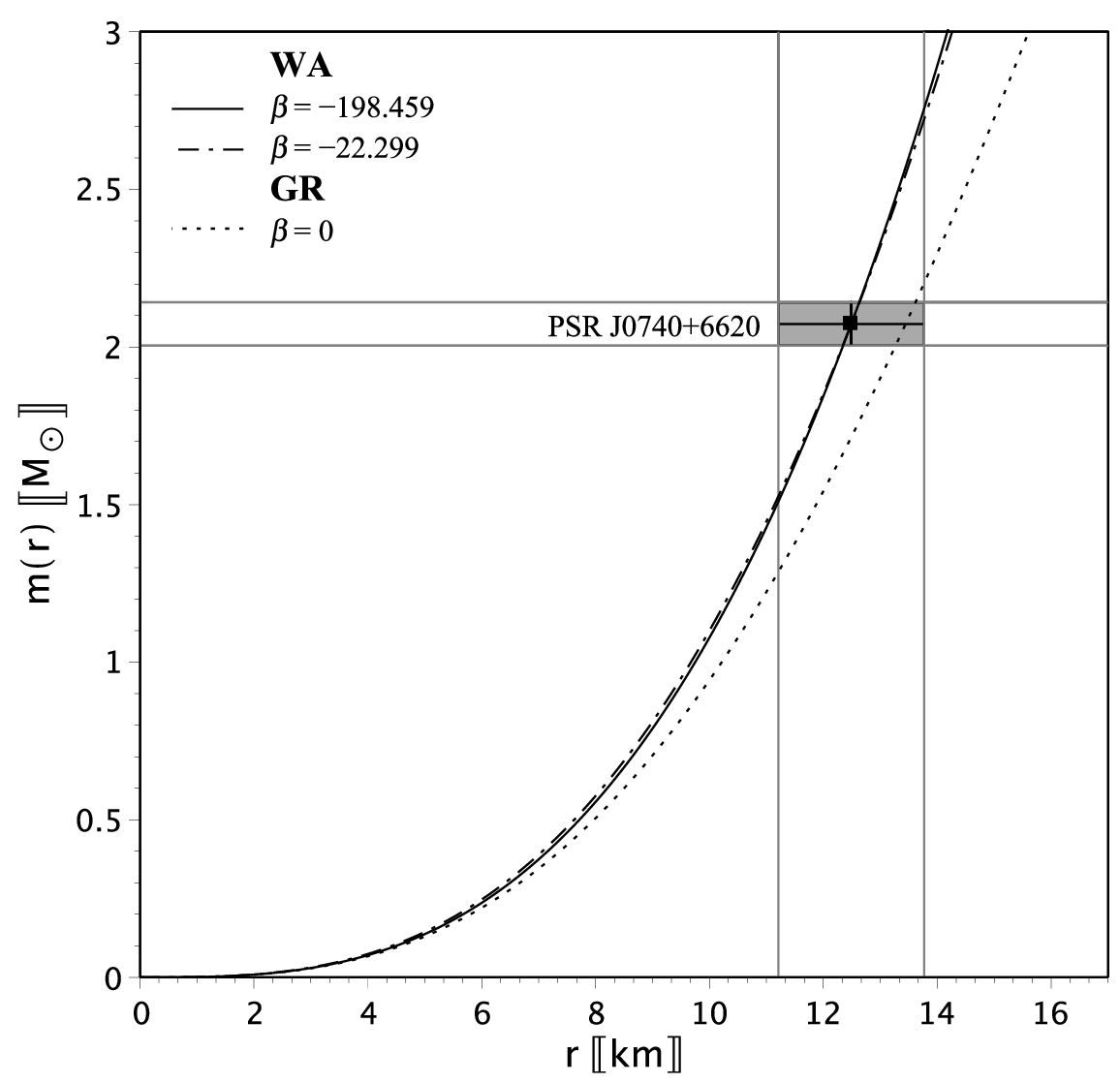}
    \caption{The astrophysical constraints on the radius and the mass of the pulsar PSR J0740+6620 from updated NICER and XMM-Newton data $\mathcal{R}=12.49^{+1.28}_{-0.88}$ km and $M=2.073^{+0.069}_{-0.069}M_\odot$ \cite{Salmi:2024aum}. The Weyl anomaly with $\beta=-198.459$ (associated to \eqref{eq:const1}) and $\beta=-22.299$ (associated to \eqref{eq:const2}) fits the observation very well better than GR ($\beta=0$).}
    \label{fig:Fig2}
\end{figure}

We note that the parameter space $\{\beta, C\}$ associated to Buchdahl limit cannot set restrictive constraints on the obtained results. Nevertheless, it shows in a clear way how Buchdahl limit modification is correlated to Weyl anomaly contribution. Therefore, we use the 3.6 Years of NICER Data of the pulsar PSR J0740+6620 and XMM-Newton data which gives $M=2.073^{+0.069}_{-0.069} M_\odot$ and $\mathcal{R}=12.49^{+1.28}_{-0.88}$ km \cite{Salmi:2024aum}. Consequently, the pulsar compactness $C=0.495 \pm 0.059$. Now by substituting into the mass function \eqref{eq:mass_fn}, we get $\beta=-198.459$ where the first set of model parameters \eqref{eq:const1} is used, and $\beta=-22.299$ where the second set of model parameters \eqref{eq:const2} is used. The plot of the mass function for both branches of Weyl anomaly solution is given by Figure \ref{fig:Fig2} in addition to the GR case $\beta=0$. The figure shows clearly that the mass curve fits the observational constraints better when Weyl anomaly is considered. Recalling Fig. \ref{fig:fig1a} which shows the parameter space $\{\beta, C\}$ associated to Buchdahl limit, we find that the pulsar PSR J0740+6620 with $C\sim 0.5$ and $\beta\sim -22$ or $\beta \sim -198$ lies in the feasible region satisfying Buchdahl inequality.

\begin{figure*}
\centering
\subfigure[MR diagram ($\beta=-198.459$)]{\label{fig:beta1}\includegraphics[width=0.32\linewidth]{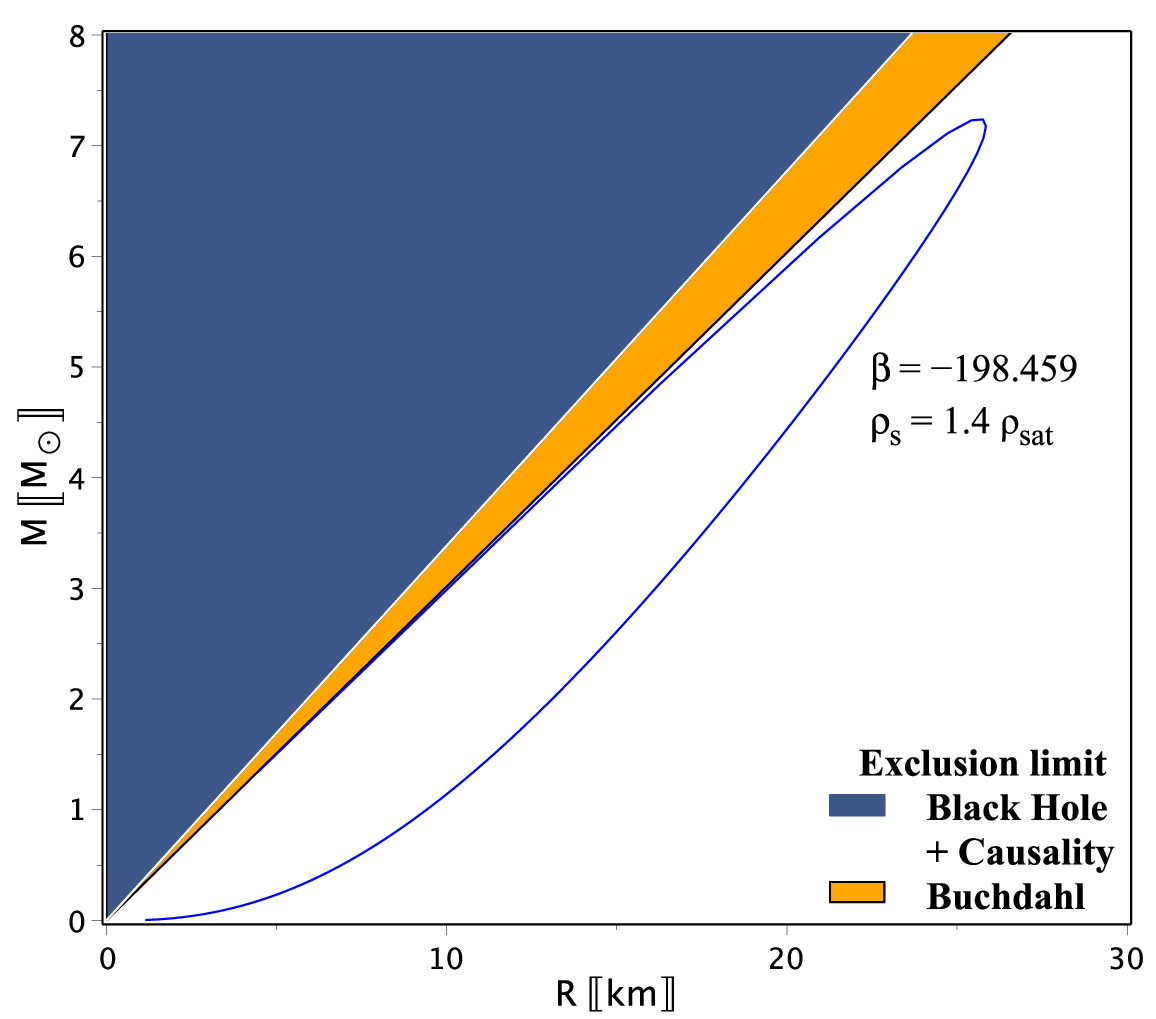}}\hspace{0.2cm}
\subfigure[MR diagram ($\beta=-22.299$)]{\label{fig:beta2}\includegraphics[width=0.32\linewidth]{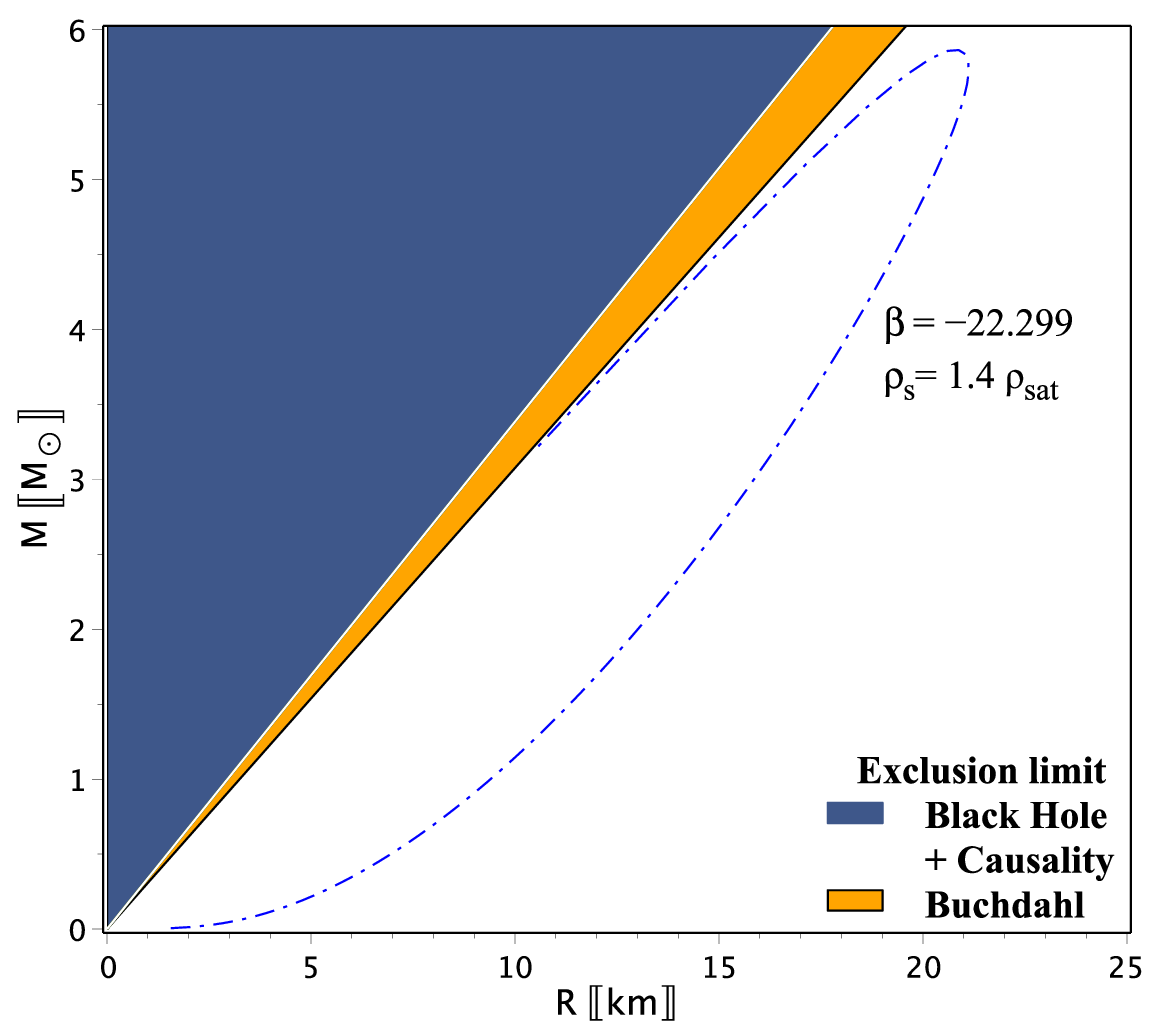}}\hspace{0.2cm}
\subfigure[Observational constraints]{\label{fig:NICER}\includegraphics[width=0.32\linewidth]{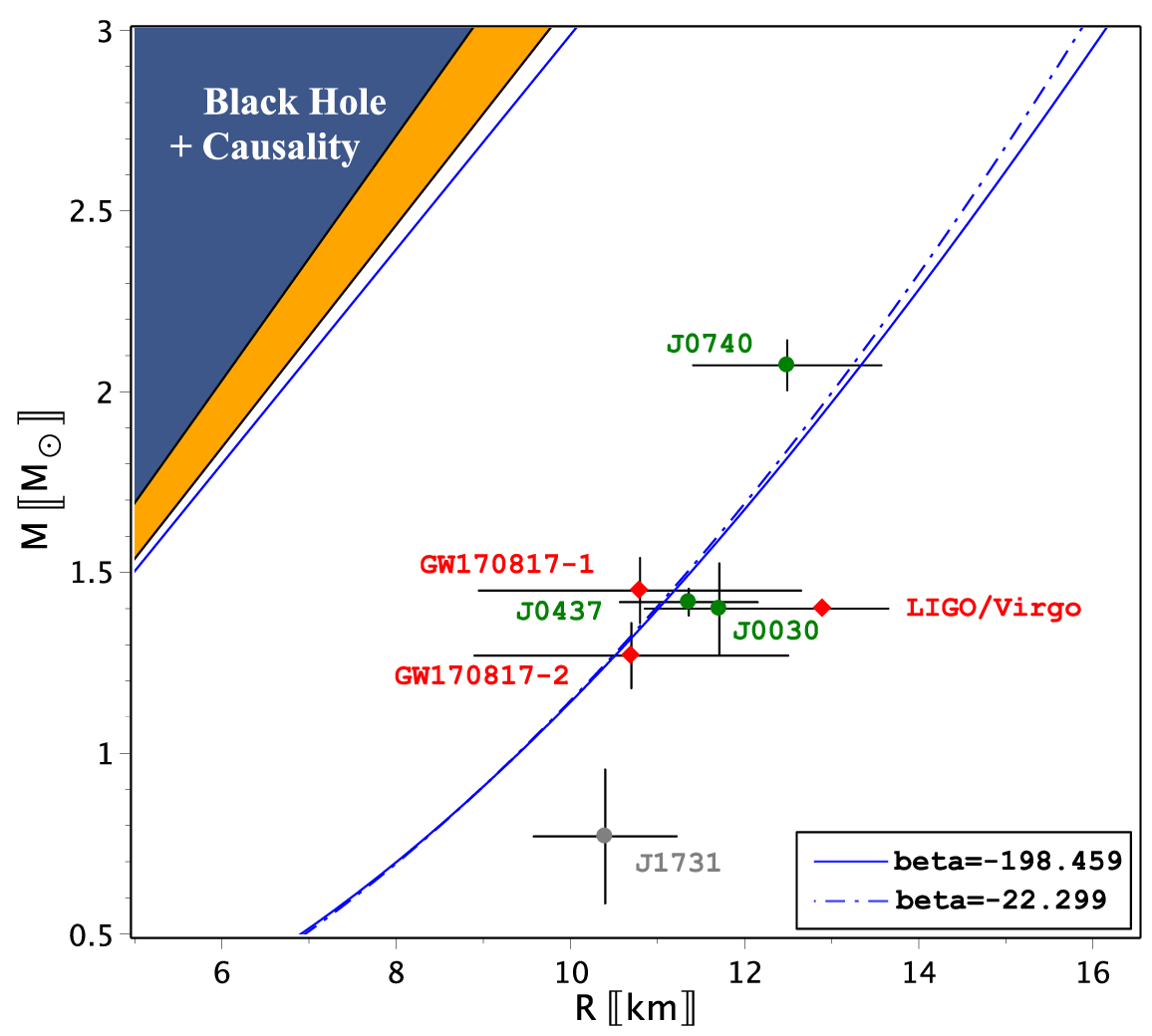}}
\caption{MR diagram. We use the boundary condition that the surface density $\rho(r=\mathcal{R})=\rho_\text{S}=1.4 \rho_\text{sat}$, where the exclusion limits due to different physical conditions are given by shaded regions. \subref{fig:beta1} For $\beta=-198.459$, the curve does not cross Buchdahl limit where the maximum mass $M=7.24~M_\odot$. \subref{fig:beta2} For $\beta=-22.299$, the curve does not cross Buchdahl limit where the maximum mass $M=5.86~M_\odot$. \subref{fig:NICER} The MR curves for $\beta=-198.459$ and $\beta=-22.299$ with observational constraints from NICER and LIGO/Virgo.}
\label{Fig:Fig3}
\end{figure*}

The MR diagram is a powerful tool to confront compact star models with different astrophysical observations as well as theoretical constraints such as BH, causality, Buchdahl limits. We note that both BH ($C\to 1$) and causality ($v_r/c \to 1$) constraints in the present study are identical as shown in Fig. \ref{Fig:Fig3}. We use the following list of the most updated observational data on the pulsar masses (pulsar timing array) and radii (NICER+XMM-Newton+LIGO/Virgo): PSR J0740+6620, as observed by NICER and XMM-Newton data, $M=2.073^{+0.069}_{-0.069}$ and $\mathcal{R}=12.49^{+1.28}_{-0.88}$ km \cite{Salmi:2024aum}. PSR J0030+0451, $M=1.40^{+0.13}_{-0.12}M_\odot$ and $\mathcal{R}=11.71^{+0.88}_{-0.83}$ km, using ST+PDT NICER and XMM-Newton \cite{Vinciguerra:2023qxq}. PSR J0437-4715, $M=1.418 \pm 0.037 M_\odot$ and $\mathcal{R}=11.36^{+0.95}_{-0.63}$ km \cite{Choudhury:2024xbk}. GW170817-1, $M=1.45 \pm 0.09 M_\odot$ and $\mathcal{R}=10.8^{+2.0}_{-1.7}$ km \cite{LIGOScientific:2018cki}. GW170817-2, $M=1.27 \pm 0.09 M_\odot$ and $\mathcal{R}=10.7^{+2.1}_{-1.5}$ km \cite{LIGOScientific:2018cki}. HESS J1731-347, $M=0.77^{+0.20}_{-0.17} M_\odot$ and $\mathcal{R}=10.4^{+0.86}_{-0.78}$ km \cite{Horvath:2023uwl}.

We plot the MR diagram for both branches $\beta=-198.459$ and $\beta=-22.299$ as given by Figures \ref{fig:beta1} and \ref{fig:beta2}, respectively. We use the boundary condition that the surface density $\rho(r=\mathcal{R})=\rho_\text{S}=1.4 \rho_\text{sat}$ which seems to fit the data, see Figure \ref{fig:NICER}. Clearly, the MR curves cannot fit the low-mass pulsar HESS J1731-347. This may point out considering non-minimal coupling between matter and Weyl anomaly. In effect, this scenario imposes extra force in the TOV equation of hydrostatic equilibrium similar to quadratic Rastall gravity case which gives a perfect agreement with observations at all mass ranges \cite{ElHanafy:2024cti}.


\section{Summary and Conclusion}\label{Sec:summary}

Neutron stars provide extraordinary laboratories to test matter at the highest densities in the universe. They are in position at the intersection of general relativity and quantum mechanics. Understanding how these two theories interact in such extreme conditions could provide insights into a unified theory of quantum gravity. While NSs are not quite at the Planck scale, at which quantum gravity shows up, the conditions inside them are much closer to it than any astrophysical object. In this sense, it is important to investigate possible impact of quantum effects on Einstein field equations at the semi-classical level. At finite temperature and baryon chemical potential, as in the NS case, the trace anomaly can be decomposed into the vacuum and the matter parts. Trace anomaly due to the matter contribution only has been investigated inside isotropic NSs as signature of conformality. The vacuum part of the trace anomaly can be expressed in terms of the vacuum expectation value of the energy-momentum tensor at zero temperature and zero baryon chemical potential. Therefore, we consider the role of quantum effects of this part due to Weyl ``trace" anomaly on compact star interiors. Weyl anomaly is one of the interesting phenomena in quantum field theory on curved spaces that arises at one-loop level for a collection of conformal fields.

We assume the following to setup the present model: (i) the spacetime interior/exterior is described by a static spherically symmetric metric. (ii) the fluid inside the compact star is anisotropic as expected in high dense medium. (iii) conformal flat metric is imposed, i.e. vanishing Weyl tensor, which allows to write the vacuum trace anomaly in terms of Euler density. (iv) At finite temperature and chemical potential, trace anomaly can be decomposed into matter and vacuum parts. We assume these parts to cancel each other in order to obtain an exact solution. We ignore the corrections, due to the non-commutation of regularization and taking the trace operations, in non-conformal theories as they are expected to be small.

We derive a non-singular exact interior solution of anisotropic compact star in presence of Weyl anomaly. We show that the matter and the vacuum sectors, each holds conservation equation. This allows to write the TOV equation of hydrostatic equilibrium in presence of Weyl anomaly. Additionally, by applying the appropriate boundary conditions at the star surface, we show that the model parameters can be expressed in terms of the compactness of the star, $C$, and another parameter, $\beta$, which characterizes Weyl anomaly contribution. We show that two families of solutions can be obtained.

We investigate possible modification to the classical Buchdahl limit on the maximum compactness. We plot the parameter space $\{\beta, C\}$ showing feasible regions which describe the modifications to Buchdahl inequality for different $\beta$ values. Excluding positive $\beta$ regions as constrained by the regularization scheme, we find that the compactness reaches the BH limit $C\to 1$ where $\beta\to -1$, where the classical Buchdahl limit, $C<8/9$, is recovered at $\beta=0$ (the GR case) and at large $|\beta|$.

We use the recently measured mass and radius of the Pulsar PSR J0740+6620 to constrain the model parameters. We use the 3.6 Years of NICER + XMM-Newton data of the pulsar PSR J0740+6620 which gives $M=2.073^{+0.069}_{-0.069} M_\odot$ and $\mathcal{R}=12.49^{+1.28}_{-0.88}$ km \cite{Salmi:2024aum}; consequently the pulsar compactness $C=0.495 \pm 0.059$. This enables us to determine the values of Weyl anomaly parameter $\beta=-198.459$ and $\beta=-22.299$ according to the solution family used. Finally, we plot the corresponding MR diagram which shows the capability of the model to describe very heavy compact objects beyond the lower mass-gap. We find that the obtained MR curves for both $\beta$ values fit the observational data from NICER + XMM-Newton + LIGO/Virgo very well by choosing a surface density 1.4 times the nuclear saturation density. However, much lower value is needed to fit the low-mass pulsar HESS J1731-347. This may indicate the need to assume the case when matter and Weyl anomaly are non-minimally coupled, similar to quadratic Rastall gravity which gives a perfect agreement with observations at all mass ranges.

\end{document}